# Optothermal rotation of micro-/nano-objects in liquids

**Authors:** Hongru Ding,[1] Zhihan Chen,[2] Carolina Ponce,[1] and Yuebing Zheng[1,2,*]

**Affiliations:**

[1]Walker Department of Mechanical Engineering, The University of Texas at Austin, Austin, TX 78712, USA.

[2]Materials Science & Engineering Program and Texas Materials Institute, The University of Texas at Austin, Austin, TX 78712, USA.

*Correspondence to: zheng@austin.utexas.edu.

## Abstract

Controllable rotation of micro-/nano-objects provides tremendous opportunities for cellular biology, three-dimensional (3D) imaging, and micro/nanorobotics. Among different rotation techniques, optical rotation is particularly attractive due to its contactless and fuel-free operation. However, optical rotation precision is typically impaired by the intrinsic optical heating of the target objects. Optothermal rotation, which harnesses light-modulated thermal effects, features simpler optics, lower operational power, and higher applicability to various objects. In this Feature Article, we discuss the recent progress of optothermal rotation with a focus on work from our research group. We categorize the various rotation techniques based on distinct physical mechanisms, including thermophoresis, thermoelectricity, thermo-electrokinetics, thermo-osmosis, thermal convection, and thermo-capillarity. Benefiting from the different rotation modes (i.e., in-plane and out-of-plane rotation), diverse applications in single-cell mechanics, 3D bio-imaging, and micro/nanomotors are demonstrated. We conclude the article with our perspectives on the operating guidelines, existing challenges, and future directions of optothermal rotation.

## Introduction

Awarded the Nobel Prize in Physics 2018, optical tweezers have proved to be an effective instrument for non-contact spatial manipulation of micro/nanoparticles as well as living biological samples.[1-8] On the basis of optical tweezers, various optical rotation platforms have been developed for exquisite control of micro/nanoscale targets over their rotational degrees of freedom using light (e.g., laser beams).[9-11] The light-powered rotation, termed optical rotation, holds significant application potential in nano-torque sensing,[12-14] nanosurgery,[15, 16] micro/nanofluidic systems,[17, 18] and so on. To achieve optical rotation, dynamic intensity profile,[19, 20] nonlinear polarization,[12, 21, 22] optical angular momenta,[23, 24] or radiation pressure[25] is typically utilized to produce asymmetric light-matter interactions for the generation of optical torques. In addition, sophisticated designs are also required on the rotating targets, which typically possess on-demand geometry, optical birefringence, or specific compositions.[26-29] The rigorous requirements for light beams and properties of particles largely restrict the broader applications of optical rotation. Moreover, the optical rotation of nanoparticles is extremely challenging as optical forces and torques decrease substantially with particle size.[6] Though high-power light beams can be used for rotation at the nanoscale level, the resultant optical heating will lead to strong Brownian motions that significantly reduce the frequency stability of the rotation.[30]

To overcome the bottlenecks of complex optics, limited applicability, and heat-induced instability of optical rotation, heat-mediated optical rotation, i.e., optothermal rotation, has been developed for rotary control of micro-/nano-objects by leveraging optical heating to generate thermal forces and torques. Over the past two decades, researchers have established multiple theories to describe the translational migration of colloids and living objects under temperature gradient fields.[31-36] Consequently, different optothermal manipulation techniques have been developed,[37-52] leading to applications in nanofabrication,[53-58] chemical and biological sensing,[59-61] and cargo delivery.[62, 63] Based on this, optothermal rotation has been further developed recently for precise and versatile rotations of colloidal particles and living cells with striking advantages such as low operation power, high applicability, and long working distance. Through the synergy of diverse light-induced thermal forces and/or optical forces, researchers, including us, manage to rotate objects of diverse sizes, materials, and shapes with low-power laser beams and simple optics for the growing demands in sophisticated biological measurements, imaging, and the development of nano-engines. This Feature Article focuses on the development of the optothermal rotation of micro/nanoscale objects in liquids (Fig. 1). We first introduce the fundamentals of different thermal forces that can drive stable optothermal rotation. Then, different approaches to achieving in-plane and out-of-plane rotation in respect of the substrates are reviewed based on their working mechanisms and various potential applications. Finally, we provide our perspectives on existing challenges and future directions of optothermal rotation.

## 2. Fundamental mechanisms

Different physical mechanisms for optothermal rotation in liquids are introduced in this section, including thermophoresis, thermoelectricity, thermo-electrokinetics, thermo-osmosis, thermal convection, and thermo-capillarity. The thermal forces originating from those optothermal phenomena are employed to control the translational movements of the targets and drive their stable rotation.

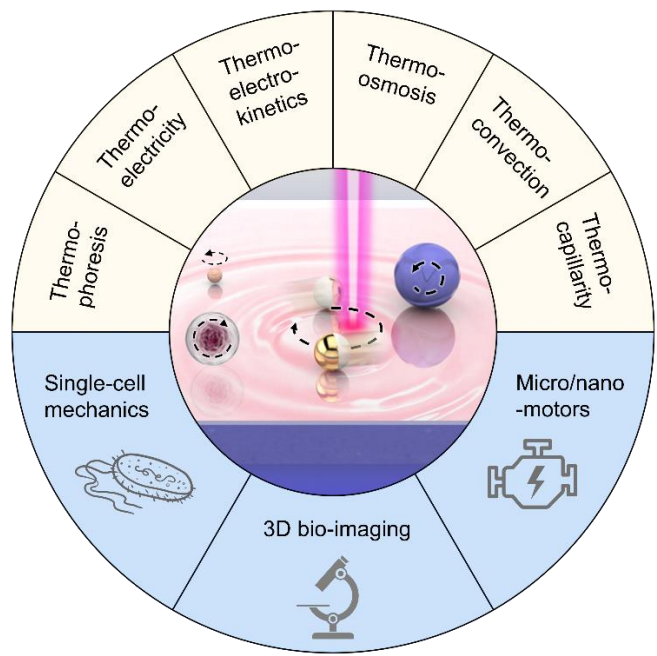

Fig. 1 Overview of optothermal rotation. Orbiting, in-plane and out-of-plane spinning of different colloids and living objects can stem from a variety of optothermal phenomena

including thermophoresis, thermoelectricity, thermo-electrokinetics, thermo-osmosis, thermal convection, and thermo-capillarity. Optothermal rotation is promising in opening new paths toward single-cell mechanics, 3D bio-imaging, and micro/nanomotors.

## 2.1. Thermophoresis

Thermophoresis (also known as thermodiffusion or the Soret effect) is defined as the directed migration of tiny particles (e.g., ions, molecules, and colloidal particles) along a temperature gradient in liquids.[64-67] The temperature gradient functions as a general force that directs the suspended particles to a cold or warm region at a certain velocity given by

$$\boldsymbol{u} = -D_T \nabla T \qquad (1)$$

where $D_T$ is the thermophoretic mobility and $\nabla t$ denotes the temperature gradient. in a nonuniform temperature field, the movement of the particles is also affected by Brownian diffusion which competes with thermodiffusion. Thus, Soret coefficient ($S_T = D_T/D$) is proposed to describe the thermophoretic migration in a general manner. With $S_T > 0$ (OR $S_T < 0$), the suspended particles show thermophobicity (or thermophilicity) and move toward the cold (or hot) region. In addition, a large $S_T$ in magnitude means that the thermophoresis dominates over the Brownian motion, indicating that the particle can have more directional motion under temperature gradient fields.

## 2.2. Thermoelectricity

Liquid thermoelectricity describes the generation of an electric field, i.e., thermoelectric field, from the charge separation of ions in electrolyte solutions under a temperature field.[38] In a thermal equilibrium state, cations couple with anions due to electrostatic interactions. While in a non-equilibrium state, a thermal gradient drives cations and anions with different speeds and directions, depending on the ions' size and solvation energy.[68] Additionally, the diffusion of molecules with high polarity (e.g., water) under a certain temperature can also lead to nanoscale separation of atoms with different electronegativity.[69] These spatially separated charges (ions and atoms with partial charges) lead to a bulk thermoelectric field given by

$$E_{TE} = \frac{\int e(n_+ - n_-)\,\mathrm{d}z}{\varepsilon} \qquad (2)$$

where $n_+$ and $n_-$ are volumetric number densities of cations and anions, respectively. $\varepsilon$ denotes the solvent permittivity. Since most colloids have surface charges, a thermoelectric field can be employed to regulate the translational and rotational motions of the colloids through electrostatic interactions.

### 2.3 Thermo-electrokinetics

In analogy with optical electrokinetics, where the migration of colloids is induced by the non-uniform electric field based on a photoconductive substrate,[70, 71] thermo-electrokinetics describes a similar electrokinetic (EK) migration that stems from the electric fields from the temperature-responsive substrates.[44, 72] Most colloids are negatively charged due to their ionized acid groups on the surface.[31] Similarly, substrates can carry surface charges by coating a thin layer of acid groups (e.g., carboxylic acids). Particularly, the decrease of temperature generally promotes the dissociation of the acid molecules (e.g., COO$^-$ and H$^+$) according to the Van't Hoff equation[73, 74]. Therefore, the surface-charge gradients of colloids and substrates can be tuned by the temperature field, which leads to an EK torque that can power the rotation of the colloids. The EK force is given by

$$\boldsymbol{F}_{\text{EK}} = \oint \sigma_r dA_r \iint \boldsymbol{E}_{\parallel} dx dy \tag{3}$$

where $\sigma_T$ is the surface charge density of the colloidal particle, $dA_r$ is the differential area element on the particle surface, and $\boldsymbol{E}_{\parallel}$ is the parallel component of the electric field.

### 2.4 Thermo-osmosis

Thermo-osmosis describes the flow parallel with a surface under a temperature gradient, which is typically directed toward higher temperature regions. Specifically, excess hydrostatic pressure can be induced by a temperature gradient within an electric double layer near the surface. The gradient of the hydrostatic pressure is opposite to the temperature gradient and leads to a creeping flow parallel to the surface. Thus, when a colloidal particle stays in a temperature field, the force from the thermo-osmotic flow drives the particle moving in the opposite direction of the flow at the velocity[31]

$$\boldsymbol{u} = -\frac{\varepsilon \zeta^2 \nabla T}{3\eta T} \tag{4}$$

where η is the solvent viscosity, and ζ is the surface potential of the particle. Moreover, when a substrate is heated, the thermo-osmotic flow in the vicinity of the substrate can also be used to direct and rotate neighboring micro/nanoparticles.[36, 75, 76]

**2.5 Thermal convection**

Thermal convection, known as natural convection, results from buoyance forces exerting on fluids with heat-induced density variations.[77] Like the violent vibration of atoms in solids at high temperatures,[78, 79] fluid molecules scatter and separate in the vicinity of a hot spot, causing the fluid to be less dense. Due to buoyance forces, the less-dense fluid moves upward, and the cooler fluid gets denser and sinks. Meanwhile, the surrounding fluid moves toward the hot spot due to fluid continuity. Thermal convection is typically bulky and not suitable for single-particle manipulation. Therefore, optical heating upon laser illumination on light-absorbing nanostructures has been proposed to achieve localized temperature gradient fields for accurate control of thermal convection flows and delicate micro/nano-manipulation.[80, 81]

**2.6 Thermo-capillarity**

Thermo-capillary flow is one type of Marangoni effect: mass transfer along a fluid-fluid interface happens due to the gradient of surface tension. Surface tension can be tuned by concentration, temperature, and electrical potential. The cool liquid that has a high surface tension can pull the surrounding warmer liquid that has a lower surface tension. This leads to a thermo-capillary flow, whose velocity is given by

$$\boldsymbol{u} = \frac{\mathrm{d}\gamma}{\mathrm{d}T}\frac{\nabla T}{\eta} \qquad (5)$$

where γ is the interfacial surface tension. Due to its capability of rapid mass transfer, the optothermo-capillary flow at fluid-liquid interfaces near micro/nanostructures has been utilized for particle manipulation, digital fabrication, sensing, and energy harvesting.[82-87]

## 3. Optothermal rotation techniques

Different applications of optothermal rotation are demonstrated whereby rational management of light, heat, and solutions in optothermal fluidic systems. This Feature article focuses on the contributions of Zheng Research Group to the recent progress in optothermal rotation. According to the relationship between the rotation axis and substrate, optothermal rotation can be classified into in-plane optothermal rotation and out-of-plane optothermal rotation, which both lead to different types of applications and are discussed separately in this section.

**3.1. In-plane rotation**

**Opto-thermophoretic rotation techniques.** Opto-thermophoresis has been widely used in micro/nano-manipulation by tailoring temperature gradient fields and managing $S_T$ of target objects.[88] Duhr et al. demonstrated on-demand opto-thermophoretic accumulation and depletion of DNA molecules by manipulating $S_T$ with temperature and salt concentration.[89] A similar dual manipulation of synthesized nanoparticles was achieved by Weinert et al. using opto-thermophoretic flows in the vicinity of microparticles with different $S_T$.[90] Later, Cichos et al demonstrated the trapping and directed swimming of nanoparticles with thermophoretic force fields generated from the optical heating of gold micro/nanostructures.[91, 92] More recently, we developed opto-thermophoretic tweezers for dynamic and low-power manipulation of micro-/nano-objects including lipid vesicles which are challenging to be trapped by optical tweezers.[41, 48] Despite the tremendous progress in opto-thermophoretic manipulation, opto-thermophoretic rotation is still challenging.

To fill this gap, we have developed opto-thermophoretic platforms for both rotational and translational manipulation of synthesized and biological cells.[49, 93] The working principle of our platforms is illustrated in Fig. 2a. First, a real-time tailorable and reconfigurable temperature field is established upon the illumination of a laser beam on a uniform light-absorbing substrate (Fig. 2a(i)). Since the alignment of solvent molecules at the liquid-particle interface depends on the surrounding temperature (Fig. 2a(ii)), a gradient of interfacial entropy is then generated under the light-generated temperature field. According to Anderson's model, the entropy gradient causes an interfacial slip flow that exerts a thermophoretic force on the particle's surface.[94] The thermophoretic velocity is given by[32]

$$\boldsymbol{u} = -\nabla T \frac{\varepsilon}{2\eta T} \frac{2\Lambda_l}{2\Lambda_l + \Lambda_p} (1 + \frac{\partial \ln\varepsilon}{\partial \ln T})\zeta^2 \qquad (6)$$

where $\Lambda_l$ and $\Lambda_P$ are the thermal conductivities of the liquid and the particle, respectively. The permittivity term, $\partial\ln\varepsilon/\partial\ln T$, is a function of interfacial entropy and the key for the modulation of opto-thermophoretic movement. For instance, we achieved on-demand opto-thermophoretic trapping of 1-μm polystyrene (PS) sphere in water as illustrated by Fig. 2a(iii).

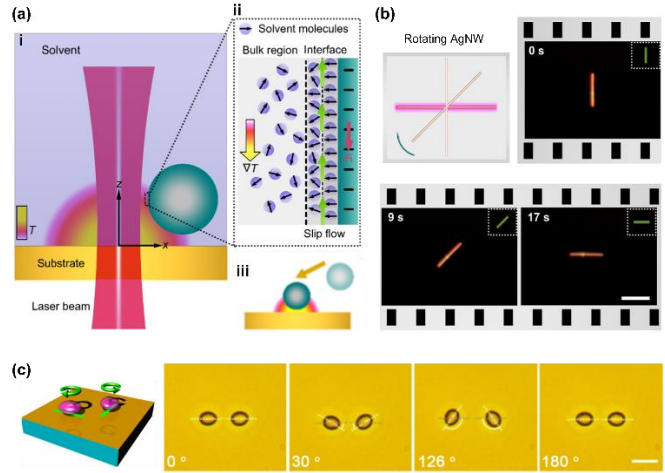

Fig. 2 Opto-thermophoretic rotation techniques. (a) Working principle of opto-thermophoretic manipulation of colloids. (b) Opto-thermophoretic rotation of a silver nanowire by a thermoplasmonic with a 532-nm laser beam. (c) Opto-thermophoretic rotation of two live yeast cells in water with two line-shaped laser beams at the wavelength of 532 nm. The light-absorb substrate is the same as (b). Scale bars: (b) 5 μm and (c) 10 μm. (a)-(b) Adapted with permission.[49] Copyright 2018, American Chemical Society. (c) Adapted with permission.[93] Copyright 2017, American Chemical Society.

The rotation of micro/nanoscale colloids can be achieved on this platform through geometric design and dynamic control of opto-thermophoretic potentials for torque generation. The optical rotation of metallic anisotropic nanoparticles has been achieved by exploiting optical aligning torque produced by laser beams with rotating linear polarization.[9] A similar opto-thermophoretic aligning torque can be achieved by reshaping and steering the laser beams. Comprehensively, a line-shape beam spot can be obtained through the digital micromirror device (DMD), which leads to a line-shape opto-thermophoretic potential. This anisotropic potential can then capture anisotropic micro-/nano-objects at the beam spot and rotate them until the status of minimum energy is reached. By rotating the line-shape beam spot, a rotary anisotropic opto-thermophoretic potential well can be obtained which continuously rotates targets through the dynamic minimum-energy status. As shown in Fig. 2b, a sliver (Ag) nanowire (12 µm × 100 nm) is rotated in isopropyl alcohol (IPA) by a rotary one-dimensional (1D) beam spot,[49] where the offset between the position of the Ag nanowire and the minimum-energy position leads to a trapping torque. In addition to synthesized particles, we have also achieved the rotation of live biological cells.[93] In water, parallel rotation of two yeast cells at an angular resolution of one degree has been demonstrated (Fig. 2c). Moreover, our platform enables the rotation of bacteria such as Escherichia coli.

Compared with other optical rotation techniques, opto-thermophoretic rotation has shown its versatility in the dynamic rotation of low-dimensional objects of diverse sizes and materials using a single low-power beam. In addition, the opto-thermophoretic torque is not dictated by the refractive index contrast between the targets and surroundings, which is superior to traditional optical tweezers.

**Opto-thermoelectric rotation techniques.** To date, thermoelectric fields have been extensively applied to the transport of charged particles such as molecules, micelles, and colloidal particles.[34, 35, 66, 95] The thermoelectric transport of charged particles is controlled by the electric force acting on the particles. Thus, the migration behaviors of the particles can be predicted by their surface charges and the thermoelectric field generated from the thermophoretic separation of ions.[46] Nevertheless, the manipulation of colloids with opposite surface charges in one solution is challenging due to the opposite signs of $S_T$. Lately, this challenge was addressed by us with the development of opto-thermoelectric tweezers.[42] By using a solution with ionic surfactants (e.g., cetyltrimethylammonium chloride (CTAC)), different manipulations including trapping, pulling, assembling, and printing of different micro-/nano-objects with arbitrary surface charges have been achieved.[40, 47, 55, 58, 96, 97] Specifically, the CTAC surfactant can be easily adsorbed on the surface of most particles through electrostatic and hydrophobic interactions. This unifies all suspended particles with positive surface charges. Meanwhile, a thermoelectric field pointing to the hot region is built due to the thermophoretic separation of CTAC micelles (CTA+) from anions (Cl-), which applies thermoelectric forces to the particles. 99

We have further extended the concept of opto-thermoelectric tweezers to an opto-thermoelectric rotation.[98] To achieve the rotation, a self-sustained thermoelectric field should be established on target particles based on their non-uniform light-absorbing structures (e.g., particles with half-coated metal). Due to the significant difference in light absorption between the metal cap and the intrinsic part, a non-uniform temperature field is established in the vicinity of the particle under illumination (Fig. 3b). Accordingly, a thermoelectric field is produced from the separation between CTAC micelles and Cl- ions around the particle, which then impose the thermoelectric force to the particle. The interplay among the thermoelectric force, optical force, and Stokes drag force (Fig. 3c) leads to a torque that drives the orbiting movements of the particle (Figs. 3d-3e). A high rotation rate of 80 rpm can be obtained by simply increasing the optical power to 3.2 mW for a higher thermoelectric field (Fig. 3f).

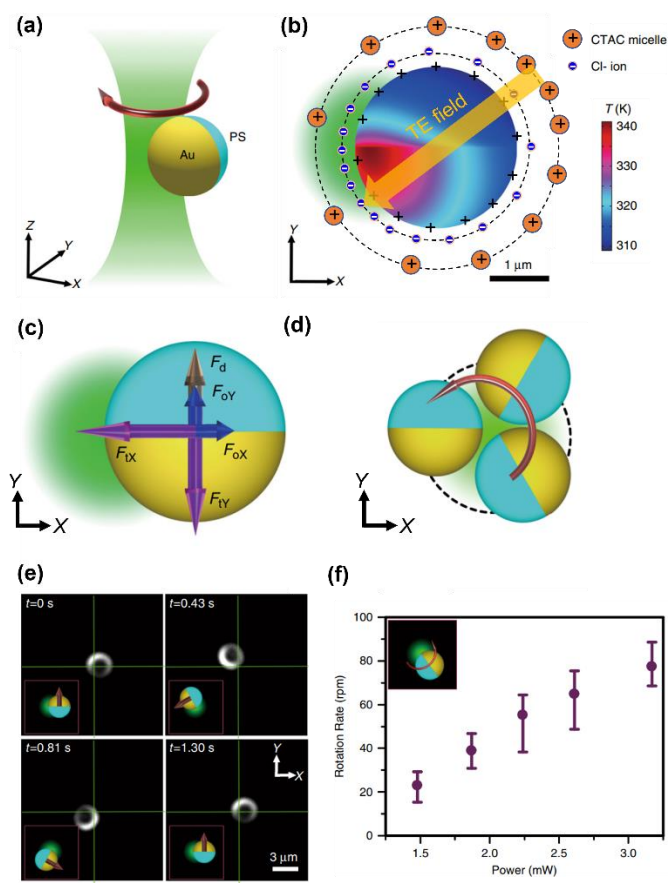

Fig. 3 Opto-thermoelectric rotation technique. (a) 3D schematic illustration of an orbiting metal-coated Janus particle driven by a laser beam. (b) Simulated temperature distribution

of a 2.7-µm Au/PS Janus particle under the illumination of a 532-nm beam at a power of 2mW. The distribution of ions and established thermoelectric field are illustrated as well. The orange and blue solid circles represent the CTAC micelles and Cl- ions, respectively. The orange arrow denotes the direction of the thermoelectric field. (c) Schematics of the thermoelectric force (purple), optical force (blue), and drag force (brown) acting on the Janus particle. (d) Two-dimensional (2D) schematics of the orbital movements of the particle powered by the beam. (e) Time-resolved dark-field images of the rotation of the Janus particle. (f) The measured rotation rate as a function of the optical power of the beam. Adapted with permission.[98] Copyright 2020, Springer Nature.

Furthermore, opto-thermoelectric swimmers are developed on the same platform, where rotary control with thermoelectric fields is utilized to achieve delicate control of swimming directions. Specifically, if the laser beam is switched for particle rotation to another loosely focused beam, the thermoelectric torque can be turned into a thermoelectric force pointing from the PS hemisphere to the au hemisphere to induce the translational motion of the particle. Through the on-demand switch between two lasers, actively navigated swimming can be demonstrated by alternately changing the propulsion and rotation states of the particles with a feedback control system. With the capability to control particles over all degrees of freedom in an automatic manner, opto-thermoelectric swimmers are foreseen to open novel horizons in micro/nanorobotics.

**Opto-thermocapillary rotation techniques.** Marangoni flow has fuelled the development of tiny motors ranging from nanometers to centimeters in the last two decades.[99, 100] Marangoni flow can be generated under external fields such as concentration and temperature fields. Particularly, optical-heating-induced Marangoni flow, i.e., opto-thermocapillary flow, has recently emerged as an effective solution for surfactant-free and high-spatiotemporal-precision micro/nanomanipulation, including rotary control of micro-/nano-objects.[82, 101-103] As the thermocapillary flow typically occurs at the liquid-fluid interface, the intrinsic Brownian motion will disturb the precise manipulation of objects, especially for nano-objects below 100 nm.

Recently, we have achieved stable and fast in-plane rotation of sub-100 nm light-absorbing particles via opto-thermocapillary forces (Fig. 4a).[104] The key to opto-thermocapillary manipulation is the thin layer of polymers on the substrate, which can be selectively turned into a quasi-liquid phase upon laser irradiation and lead to the thermocapillary flow.[105] In this work, a solid layer of CTAC is placed between a glass substrate and 80-nm gold nanoparticles (AuNPs). Upon the illumination of a 660-nm laser beam on the AuNP, a localized phase transition of CTAC is triggered by the optical heating of the particle. As shown in Fig. 4b (top), a maximum temperature of 593.4 K is obtained on the illuminated AuNP at the optical power of 10 mW. A very thin layer of CTAC in the vicinity of the AuNP (<15 nm) is then heated above its phase transition temperature (505-510 K). Therefore, a localized liquid-air interface is formed around the illuminated particle. Owing to the multifaceted asymmetry of the AuNP, a temperature gradient exists at the liquid/air interface contacting the AuNP. Meanwhile, thermocapillary stress was generated on the particle's surface because of the interfacial surface tension gradient. With $d\gamma/dT \sim -0.097$ $mNm^{-1}K^{-1}$. A considerable tangential opto-thermocapillary force can then be obtained to trigger an in-plane rotation (Fig. 4c), while the synergy of capillary forces and optical forces maintain the stable orbital movements (Fig. 4d). Figure 4e shows the real-time rotation of an 80-nm AuNP under laser excitation, whose statistic displacement results (Fig. 4f) indicate that the AuNP rotates stably in a circular orbit about the laser beam.

This opto-thermocapillary rotation platform may open a novel path toward nanomachines. Contrary to other existing light-powered machines, our nanomachines have suppressed Brownian motion and could serve as fuel-free and gear-free engines to convert optical energy into mechanical work for various nano-electro-mechanical systems.

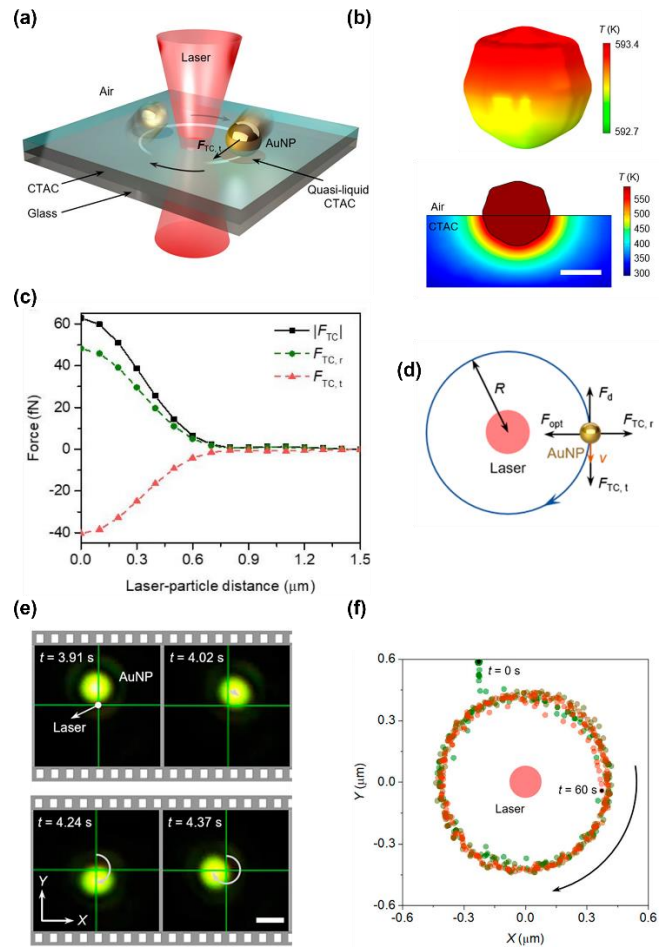

Fig. 4 Opto-thermocapillary rotation technique. (a) 3D schematic illustration of an orbiting light-absorbing nanoparticle driven by a laser beam. (b) Top: Simulated 3D temperature distribution of an 80-nm AuNP under the illumination of a 660-nm beam at a power of 10mW. Bottom: Simulated 2D temperature distribution of CTAC film near the AuNP. Scale bar: 50 nm. (c) Simulated thermocapillary force (black squares) acting on the AuNP. The tangential and radial components of the force are denoted by red triangles and green circles, respectively. (d) Schematics of the thermocapillary force, optical force, and drag force acting on the AuNP particle. (e) Time-resolved dark-field optical images of an orbiting 80

nm AuNP. Optical power: 6 mW. Scale bar: 1 μm. (f) Centroid tracking of the rotating AuNP. Adapted with permission.[104] Copyright 2022, American Chemical Society.

### 3.2. Out-of-plane rotation

Optothermal rotation techniques that enable the out-of-plane micro/nanorotation (i.e., rotation of an object around an axis parallel to the substrate) are featured in this section. Compared to in-plane rotation, out-of-plane rotation finds a variety of unique applications in fields such as single-cell mechanics, 3D bio-imaging, and micro-/nano-surgery.[16, 106-112] However, out-of-plane rotation is known to be more challenging to accomplish especially for highly symmetric or isotropic targets.[113] In the past five years, a few groups have proposed diverse strategies for out-of-plane rotation under broken symmetries generated in light-driven temperature fields.[44, 72, 75, 114-117]

**Opto-thermo-electrokinetic rotation techniques.** Historically, the main issue in achieving light-driven out-of-plane rotation has centered in generating torque about an axis perpendicular to the optical axis for rotation while simultaneously forming a trap well to overcome Brownian motion for stable rotation. It is inherently challenging to produce perpendicular optical torques despite the convenience of the generation of parallel optical torques through different strategies.[9] Light-driven out-of-plane rotation has been observed occasionally,[115] however, the maintenance of a stable rotation at a target location is challenging. Moreover, it becomes more difficult to break symmetries for the rotation of homogenous and perfectly symmetric spheres.[12, 28] Recently, Xie et al. developed a dual-beam optical tweezers system enabling the out-of-plane rotation of various cells for microsurgeries.[118-120] Nevertheless, its reliance on two high-power beams restricts its application within the out-of-plane rotation of nano-objects, especially for ones of subwavelengths. Besides, the direct illumination of the two intense beams may also cause photodamage to delicate cells.

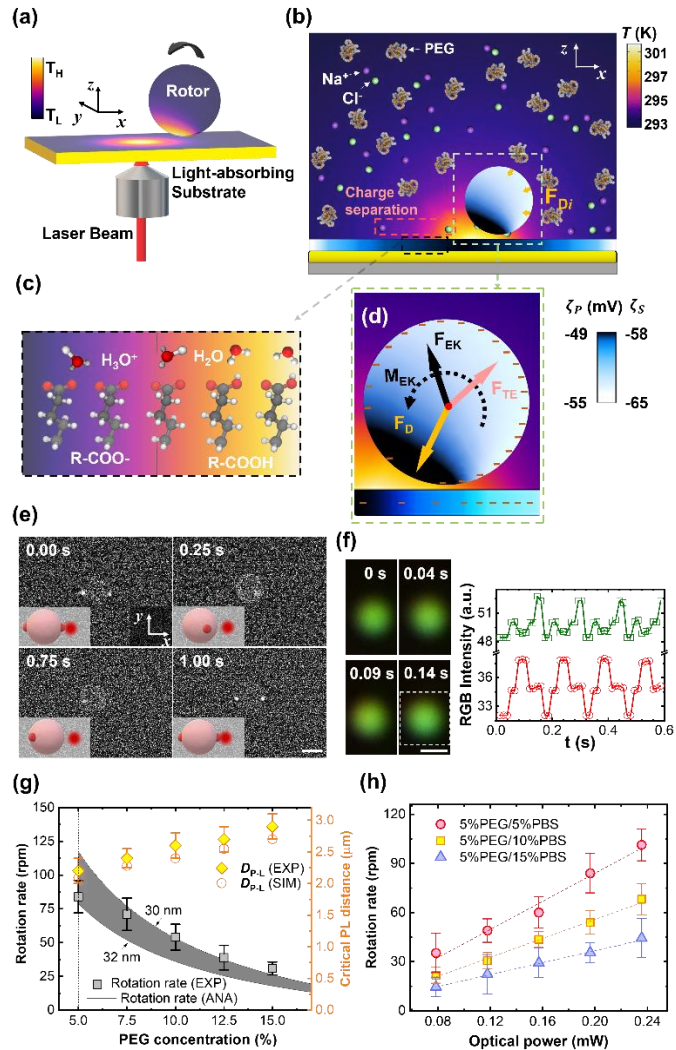

Fig. 5 Opto-thermo-electrokinetic rotation technique. (a) 3D schematic illustration of the experimental setup. (b) Simulated 2D temperature distributions of the rotor, substrate, and the surrounding liquid. Yellow arrows indicate discrete depletion forces on the rotor. (c) The temperature effect on the dissociation of carboxylic function groups. (d) Schematics of optothermal forces and torque on the rotor. The yellow, red, and black arrows denote the depletion, thermoelectric, and thermo-electrokinetic forces, respectively. A net torque (black curved arrow) can be generated on the particle at a certain position where a balance is reached among the three forces. The temperature-dependent distributions of negative charges on the surface of the particle and substrate are indicated by the "−" symbols. (e) Time-resolved fluorescent images of out-of-plane rotary 2.8-μm PS particle. Optical power: 78.4 μW. Scale bar: 2 μm.

(f) Left: Real-time dark-field optical images of a rotating 300-nm PS/Au Janus particle. Right: Red-green-blue (RGB) intensity extracted from the dark-field optical images of the Janus particle. The white dash rectangle in the left figure marks the selected area from which the RGB intensity is recorded. (g) Rotation rate and laser-particle distance ($D_{P-L}$) of a 2.8-μm PS particle versus PEG concentration in 5% PBS solutions at an optical power of 196 μW. The grey squares and black lines are experimentally measured (indicated as "EXP") and analytic (indicated as "ANA") rotation rates, respectively. The grey region represents the rotation rates calculated as the particle-substrate gap ranges from 30 to 32 nm. The yellow diamonds and circles are measured and simulated (indicated as "SIM") $D_{P-L}$, respectively. The dashed line marks the threshold PEG concentration (5%), where the rotation rate reaches a maximum value. (h) The measured rotation rate of the PS particle in 5% PEG solutions with different PBS concentrations versus the optical power of the laser beam. The dashed lines are linear fittings of the measured values. Adapted with permission.[72] Copyright 2022, American Association for the Advancement of Science.

To overcome these obstacles, we propose a universal rotation strategy based on opto-thermo-electrokinetic coupling, which enables the out-of-plane rotation of particles with diverse sizes, materials, and shapes using low-power and simple optics.[72] The working principle of opto-thermo-electrokinetic rotation is illustrated in Fig. 5a. The particles and substrate are immersed in a solution containing polyethylene glycol (PEG) and phosphate-buffered saline (PBS). A single laser beam is used to create a local temperature field between the charged particle and a light-absorbing substrate with carboxylic acid–terminated alkanethiol self-assembled monolayers. Under the temperature gradient, all the ions and molecules diffuse from the hot to the cold region, leading to the charge separation of $Na^+$ and $Cl^-$ ions, and the concentration gradient of PEG molecules (Fig. 5b). Moreover, the temperature gradient leads to the dissociation of acid functional groups on the substrate's and particle's surfaces (Fig. 5c),[73, 74] which results in the gradients of surface charge, respectively (Fig. 5d). Accordingly, the charge separation of $Na^+$ and $Cl^-$ ions leads a thermoelectric field, which exerts a repulsive thermoelectric force on the particle. Meanwhile, the concentration gradient of the PEG molecule results in an attractive depletion force to trap the particle around the laser beam.[43, 121] More importantly, the electrokinetic interaction between the surfaces of the particle and the substrate generates an accumulative force with the line of action not passing through the particle's centroid, which imposes a torque on the particle with the axis parallel to the substrate and finally drive its out-of-plane rotation. Under the illumination of lower-power 532-nm beam, a 2.8-μm PS microsphere in a 5% PEG/5% PBS solution rotates in an out-of-plane manner at the rate of 32.0 rpm (Fig. 5e). The two bright spots are fluorescent nanobeads attached to the particle for the observation of out-of-plane rotation. In contrast to other optical rotation systems where the laser directly illuminates onto the object, the laser beam in our system can be positioned away from the rotary objects to reduce the optical damage.

Our rotor platform displays general applicability to biological cells and synthetic particles of diverse materials, sizes, and shapes. For example, the out-of-plane rotation of a subwavelength PS/Au Janus has been achieved, whose rotation behavior is quantified by the real-time RGB signals from the scattered light of the particle (Fig. 5f). We have also explored the dependence of the rotation rate and $D_{P-L}$ on PEG and PBS concentrations, and optical power. In detail, the rotation rate decreases with PEG concentration while $D_{P-L}$ shows an opposite trend (Fig. 5g). In addition, the rotation rate gradually decreases as PBS concentration increases from 5% to 15%. Since thermo-electrokinetic interactions increase with temperature gradient, a higher rotation rate can be obtained using a beam with a higher power (Fig. 5h).

In short, our platform shows advantages compared to conventional optical rotation platforms in terms of simplicity, universality, and biocompatibility. The rotation of homogenous and perfectly symmetric spheres is accomplished here, which has not been achievable until now in any other light-driven rotor platforms because of their restricted demands on the asymmetric shape[12] or birefringence[28] of particles. Furthermore, out-of-plane rotation of single biological cells, which has only been implemented via a high-power dual-beam system in the past, is accomplished here using a low-power laser beam. With its simple optical setup, wide applicability, and low-power operation, our rotor platform becomes promising in various scientific research and applications.

**Opto-thermo-osmotic rotation techniques.** Thermo-osmotic flow, generated on the surfaces of the particles[122] or substrates[75] in nonuniform temperature fields, can be utilized for the manipulation of particles resting on substrates. The flow velocity is given by[123]

$$v_{\text{to}} = -\frac{1}{\eta} \int_0^\infty dz z h(z) \frac{\nabla T}{T} \tag{7}$$

where $h$ is the excess specific enthalpy in the boundary layer, which is positive for common solid substrates (e.g., glass) and generates flow towards the cold region. In contrast, some synthetic membranes show negative enthalpies that drive the flow toward the hot region. Therefore, thermo-osmotic flows can be modulated on demand by surface chemistry for particle rotation.[36] Lou et al have developed a model that predicts the opto-thermo-osmotic rotation of different colloidal particles on different substrates.[75] Later, Heidari et al. experimentally observed an orientation change of a Janus particle induced by thermo-osmotic flows from a glass slide in experiments.[115]

More recently, we report a new strategy for the out-of-plane rotation of single cells using opto-thermo-osmotic flows generated from light-absorbing substrates.[114] The experimental setup is shown in Fig. 6a. Two laser beams (785 nm and 532 nm) are focused at the same position on the substrate for simultaneous trapping and rotation of a single cell. The substrate is carefully designed by depositing a large amount of sub-100nm AuNPs on a glass slide to achieve localized temperature fields upon laser irradiation as well as obtain low surface charge for powerful thermo-osmotic flows. Meanwhile, the substrate has been designed to have high transparency for a 785-nm laser beam for a considerable trapping force. By rationally selecting an optical power intensity for the 785-nm beam, a single cell can first be stably trapped by the optical gradient force (Fig. 6b). Then, upon the illumination of the 532-nm beam, a nonuniform temperature field is established for the generation of the thermo-osmotic flow, which powers the out-of-plane rotation of the cell (Fig. 6c). The cross-section of the flow fields is shown in Fig. 6d. At the optical powers of 0.2 mW/µm$^2$ (532 nm) and 1 mW/µm$^2$ (785 nm), a rotation rate of ~1 Hz is obtained. At higher optical powers, a higher rate can be obtained due to enhanced thermo-osmotic flows (Fig. 6e). With its precise control of single-cell rotation, the opto-thermo-osmotic platform may provide an effective solution to the measurement of the ligand-receptor binding kinetics at the single-cell level (see the inset of Fig. 6a).

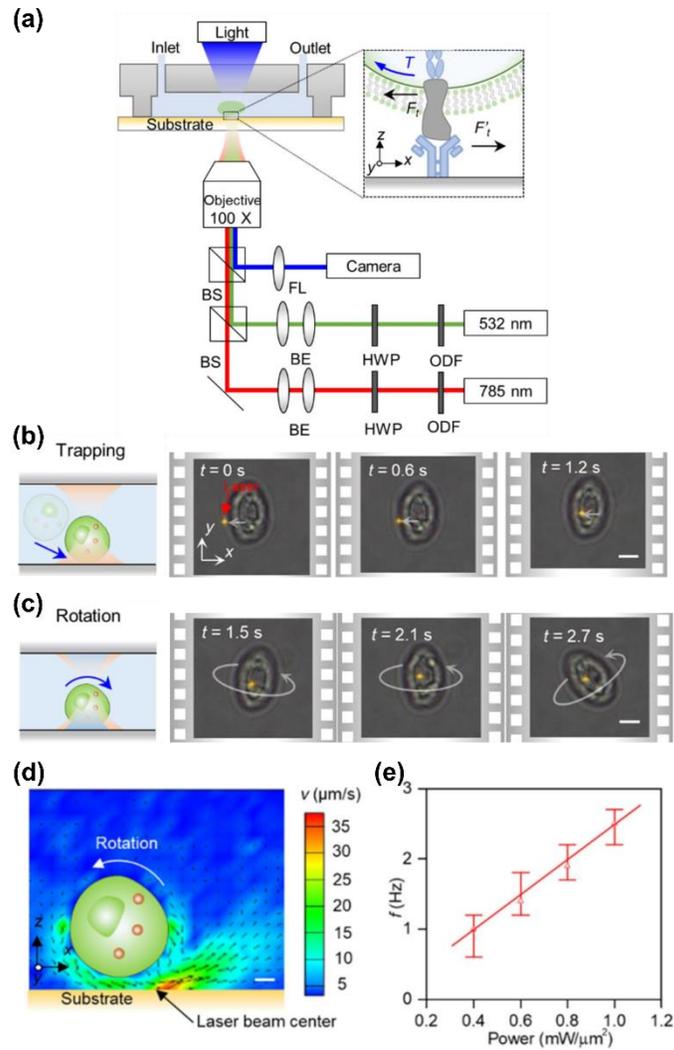

Fig. 6 Opto-thermo-osmotic rotation techniques. (a) Experimental setup. HWP: half-wave plate. ODF: optical density filter. FL: focused lens. Inset: the receptors of the rotary cell interacting with the ligand molecules immobilized on the substrate. BS: beam splitter. BE: beam expander. (b) Schematics and time-resolved optical images showing the trapping of a yeast cell (S. cerevisiae) by a 785-nm beam. (c) Schematics and time-resolved optical images showing the stable rotation of the yeast cell with the secondary beam (532 nm). Scale bars: 5 µm. (d) Cross-sectional view of the simulated flow velocity in the vicinity of the cell. Scale bar: 1 µm. (e) The dependence of the rotational rate of the trapped cell on the optical

power intensity of the 532-nm laser. The optical power intensity of the 785-nm beam is fixed at 1 mW/μm². Adapted with permission.[114] Copyright 2022, Springer Nature.

**Optothermal convective rotation techniques.** Providing delicate control of the temperature field,[81] light-generated thermal convection has been extensively exploited for the manipulation of arbitrary micro-/nano-objects, such as DNA,[124, 125] silver nanoparticles,[126] and living cells.[127] Recently, the optothermal convective flow has also been exploited for the rotational manipulation of synthesized particles and biological cells. For instance, Kumar et al. produced optothermal convective torques by heating a gold substrate with a laser beam for the out-of-plane rotation of hexagonal-shaped particles and single cells.[117] Dai et al. proposed a novel strategy for the out-of-plane rotational manipulation of hydrogel microstructures in additional to in-plane translational by the synergy of optothermal convection and opto-thermocapillary flows near a surface bubble[116] Later, they also demonstrated in-plane rotation and versatile assembly of micro-objects in different shapes based on the same convection and capillary flows stemming from surface bubbles.[128]

## 3.3 Potential applications

Featuring simple optics, low power, and wide applicability to different objects, the optothermal rotation leads to a wide range of applications including single-cell mechanics, 3D bio-imaging, and micro/nanomotors.

**Single-cell mechanics.** Rotation of single cells is important for the identification of cellular phenotypes, cell-cell communication, and, particularly, single-cell mechanics.[106, 129] Among different measurements of single-cell mechanics,[130] cell adhesion has attracted a lot of attention as it is highly involved in many biological processes such as viral infections. However, the existing single-cell adhesion kinetics measurements can only give qualitative results since they are usually performed under conditions far away from the in situ physiological environments.[131] For instance, in the study of SARS-CoV-2's entry into host cells, the difference in measured affinities of ACE2: S1 (the key parameter for infection) can be as large as one order of magnitude among different methods. For instance, the affinity measured by surface plasmon resonance assay is ~15nM[132] while that by the atomic force microscopy assay is ~120 nM[133]. In comparison, our optothermal rotation techniques show the following striking advantages: 1. Our measurement was conducted in a situation close to the real cell adhesion process, which can hardly be achieved by previous methods (Fig. 7a). 2. The measurement can be performed for diverse cells directly in complex clinical samples such as human urine without any pre-treatment (Fig. 7b). 3. Both shear affinity and tensile affinity can be measured in our platform (Fig. 7c).

**3D bio-imaging.** Enabling 3D interrogation of biological samples, contact-free rotation techniques pave new ways to non-invasive 3D bio-imaging with high resolution in all dimensions. By taking sequential images in hundreds of 2D focal planes, volumetric imaging shows high lateral resolution. However, this type of imaging technique commonly is insufficient to resolve sub-cellular events due to its low axial resolution ($> 1$ µm) caused by the missing cone along the z-axis.[134] Although advanced imaging systems such as total internal reflection fluorescence (TIRF) and selective plane illumination microscopy (SPIM) can further improve high lateral resolution, they still suffer from relatively low axial resolution and limited optical overage. In contrast, out-of-plane rotation of single cells combined with TIRF or SPIM is promising to construct 3D images of various biological samples with high resolution in all dimensions. Whyte et al. realized out-of-plane rotation of red blood cells and cancer cells with a fiber-based dual-beam laser trapping.[19] Then, they evolved it into an optofluidic device that enables single-cell rotation around an axis perpendicular to the imaging plane for 3D single-cell tomography.[135] However, all these platforms show limited throughput and are not suitable to be combined with TIRF due to the large gaps between the rotary cells and the substrates. As our optothermal rotation techniques can rotate the living cells near the substrates (Fig. 7d), our platform displays a high possibility to be integrated with TIRF and SPIM for high-resolution 3D reconstruction of various living cells. Moreover, by using digital micromirror devices or spatial light modulators, parallel rotations of multiple cells can be achieved by splitting the incident laser into multiple laser beams, which significantly improves the throughput of the 3D imaging.[93]

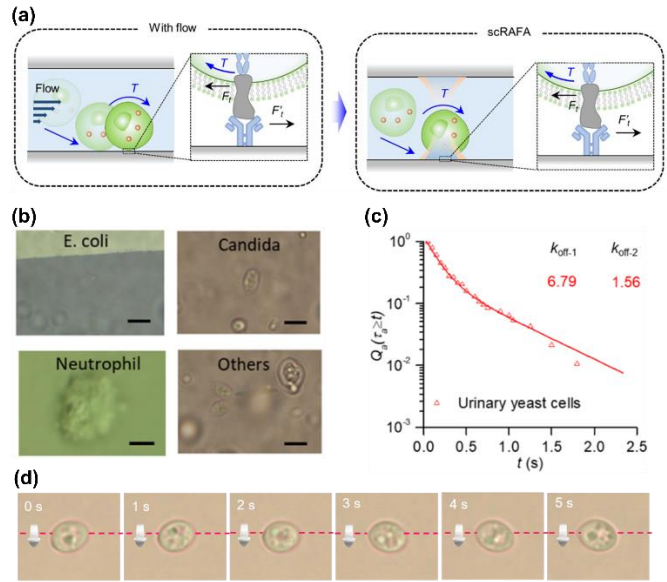

Fig. 7 Biological applications of optothermal rotation techniques. (a) Left: Schematic illustration of the adhesion process of a cell. With a flow of biofluids (e.g., blood, urine), the cell will first attach to the endothelium cells as a substrate at an inclined angle. During the rolling, the receptors on the cell and the ligands on the substrate experiences shear forces ($F_t$, $F'_t$). The curved arrow indicates the torque on the cell generated by the fluidic flow. Right: on the optothermal rotation platform, the cell suspended in liquid is first trapped on the functionalized substrate, mimicking the cell pre-attachment. Then, the measurement of the cell's shear adhesion kinetics can be obtained through the analysis of the light-driven out-of-plane rotation behavior of the cell. (b) Optical images of bacteria and cells trapped and rotated in clinical samples during the rolling adhesion measurement. Others: urinary organisms that we can hardly identify. Scale bars: 5 μm. (c) $Q_a$ ($\tau_a \geq t$), the fraction of chitin's transient adhesion with lifetime $\tau_a \geq t$, versus $t$ for urinary yeast cells. The dissociate constants of the adhesion event ($k_{off-1}$=6.79 s$^{-1}$ and $k_{off-1}$=1.56 s$^{-1}$) were extracted by fitting the experimental data with double exponential decay curves. (d) Time-resolved optical images of a rolling yeast cell. (a)-(c) Adapted with permission.[114] Copyright 2022, Springer Nature. (d) Adapted with permission.[136] Copyright 2020, Institute of Electrical and Electronics Engineers.

**Micro/Nanomotors.** Micro/nanomotors are micro/nanoscale devices that enable the conversion of energy into mechanical work. Optical, electric, chemical, and magnetic energies have been used to power micro/nanomotors for the applications of environmental remediation, cargo delivery, and micro/nanorobotics.[37, 137-140] Among them, motors driven by optical heating (i.e., optothermal motors) show unique advantages such as high general applicability, excellent temporal and spatial control with localized fields, and so on.[72, 104, 141, 142] However, most existing optothermal motors require immersion in a liquid environment, limiting their range of applications. For instance, liquid-state motors cannot be utilized in devices, such as hard drives, that need to work in a dry environment. Moreover, the precise control of a micro/nanomotor in liquid could be challenging due to strong Brownian motion. In comparison, operating in a solid environment, our opto-thermo-capillary nanomotors circumvent those problems and shed light on the underdeveloped field of solid-state micro/nanomotors.[104]

## 4. Summary and outlook

Diverse optothermal rotation techniques are developed based on different working mechanisms, which can be implemented on different substrates (thermally responsive or not), in different media (e.g., surfactant solutions, biological media, or solid polymers), and to different target objects (e.g., colloidal particles or living cells). They can either have in-plane or out-of-plane rotation frequency ranging from Hz to kHz, which leads to various applications. We conclude this Feature Article by proposing a general guideline for selecting suitable optothermal rotation techniques in terms of different target objects, followed by a discussion about the existing challenges, and future opportunities in this field.

### 4.1 General guideline

For the optimum performance, we provide a general guideline for the choices of proper optothermal rotation techniques for different target objects, including nanoparticles, microparticles, and biological cells.

**Nanoparticles**. The primary challenge in controlling the rotation of nanoparticles is their prominent Brownian motion. To counter this intrinsic effect, opto-thermoelectric, solid-state opto-thermocapillary, and opto-thermo-electrokinetic rotation techniques can provide strong trap wells to suppress the Brownian motion for the precise and stable rotation of nanoparticles. Specifically, opto-thermoelectric and solid-state opto-thermocapillary rotations are designed for highly anisotropic and light-absorbing nanoparticles, in which nonuniform temperature fields can be established under uniform illumination to generate rotation torques. In comparison, opto-thermo-electrokinetic rotation is more universal for the rotation of different types of nanoparticles regardless of their shapes, components, and light absorption. However, high surface charges of nanoparticles are typically required.

**Microparticles**. Due to the less prominent Brownian motion, optothermal rotation of microparticles can be achieved by all the means discussed in this article. We should note that for opto-thermophoretic rotation, asymmetry in either shape or light absorptivity of microparticles is required to produce torques for the rotation. Furthermore, for microparticles with a size larger than 10 µm, optothermal convective rotation can be the most suitable candidate as the convective flow offers larger forces and torques than other optothermal rotation techniques.

**Biological cells**. When considering the biocompatibility of different optothermal rotation techniques, we should point out that the solution components should be biologically safe, and the operational optical power should be low to reduce photodamage. Therefore, opto-thermophoretic, opto-thermo-electrokinetic, and opto-thermo-osmotic rotation techniques are ideal candidates for the rotation of biological cells at the single-cell level. The biological media typically contain lots of ions that make the suspended living cells cold-seeking under a temperature field. Accordingly, opto-thermophoretic rotation techniques are more suitable for robust cells that can survive in less-salty environments (such as yeast cells). In contrast, since the opto-thermo-electrokinetic and opto-thermo-osmotic rotation techniques can offer extra forces to counteract the thermophoretic forces, they can function in various biological medias for the stable rotation of living cells (even for some delicate mammalian cells such as leukocytes).[44]

### 4.2 Challenges and opportunities

The main challenges of the current optothermal rotation techniques are the potential thermal damage and limited rotation rates. Though the optothermal rotation can minimize phototoxicity due to the low optical power, thermal stress from optical heating could lead to the photothermal degradation of delicate objects.[143] Especially for optothermal convective and opto-thermo-capillary rotation techniques, higher optical powers are typically required compared to the other optothermal rotation techniques. In addition, it could be challenging to obtain an ultrahigh rotation rate (e.g., GHz) on an optothermal rotation platform. Intuitively, a larger rotation rate can be obtained from the higher optothermal torques powered by the higher optical power. However, the increasing temperature might reach the boiling point of the solution, where vapor bubbles can form and introduce strong Marangoni convection to suppress the optothermal torques. One possible solution to these two issues is to add a chip-size heat sink to lower the environmental temperature.[144] Within a lower environmental temperature, the target temperature difference for the generation of optothermal torques can be obtained with a lower peak temperature, which can reduce the thermal damage and avoid the formation of vapor bubbles[40, 93]

Optothermal rotation of objects down to a few nanometers or even atomic scale can be attractive to the study of virology and particle physics. Multiple optothermal effects have proven to be able to direct the migration of the particles with these small sizes.[37] To implement the rotation of these extremely tiny objects, one promising strategy is to use the femtosecond laser in optothermal rotation techniques. Specifically, a femtosecond laser pulse can generate a highly localized and steeper temperature field. Under a higher temperature gradient, a more robust trap well and a larger optothermal torque can be produced, which is promising to achieve stable rotary control of nano-/atomic-scale items. Besides, due to the high energy density of a femtosecond laser beam, it can directly heat the liquid through nonlinear interactions to establish temperature fields. Thus, all existing optothermal rotation techniques can work without the requirement of light-absorbing substrates or objects, which further broadens the applicability of optothermal rotations.

We believe that, once the above-mentioned bottlenecks are overcome, more applications can be implemented by optothermal rotation. For instance, with the capability to noninvasively rotate smaller objects such as viruses, optothermal rotation can be used in single-virus mechanics. Specifically, the rolling adhesion of SARS-CoV-2 (~0.1 µm) can be measured to help better understand the viral infection process. Another promising application is cell classification.

Distinguishing different types of cells of high similarity is extremely challenging for existing optical imaging techniques. Optothermal rotation of cells integrated into standard optical microscopy will provide a collection of multiangle images at a high efficiency and, in combination with machine-learning-enhanced image analysis, could effectively reduce the quantity of the training samples and improve classification accuracy.

## Author Contributions

H.D. and Y.Z. conceived the idea, designed the frame, and wrote the manuscript. Z.C. and C.P. assisted with the manuscript revision. Y.Z. supervised the project.

## Conflicts of interest

The authors declare no competing interests.

## Acknowledgments

H.D., Z.C., and Y.Z. acknowledge the financial support of the National Institute of General Medical Sciences of the National Institutes of Health (R01GM146962) and the National Science Foundation (NSF-ECCS-2001650). C.P. additionally acknowledges the financial support of UT Austin Undergraduate Research Fellowship.

## Notes and references